\documentclass[prb, twocolumn, superscriptaddress, 10pt]{revtex4}  
\usepackage{amssymb}
\usepackage{amsmath}
\usepackage{bbm}
\usepackage{graphicx}
\usepackage{color}

\begin{document}

\author{Philipp Werner}
\affiliation{Department of Physics, University of Fribourg, 1700 Fribourg, Switzerland}
\author{Jiajun Li}
\affiliation{Department of Physics, University of Erlangen-N\"urnberg, 91058 Erlangen, Germany}
\author{Denis Gole\v z}
\affiliation{Center for Computational Quantum Physics, Flatiron Institute, 162 Fifth Avenue, New York, NY 10010, USA}
\author{Martin Eckstein}
\affiliation{Department of Physics, University of Erlangen-N\"urnberg, 91058 Erlangen, Germany}

\title{Entropy-cooled nonequilibrium states of the Hubbard model}

\date{\today}

\hyphenation{}

\begin{abstract}
We show that the recently proposed cooling-by-doping mechanism allows to efficiently prepare interesting nonequilibrium states of the Hubbard model. Using nonequilibrium dynamical mean field theory and a particle-hole symmetric setup with dipolar excitations to full and empty bands we produce cold photo-doped Mott insulating states with a sharp Drude peak in the optical conductivity, a superconducting state in the repulsive Hubbard model with an inverted population, and $\eta$-paired states in systems with a large density of doublons and holons. The reshuffling of entropy into full and empty bands not only provides an efficient cooling mechanism, it also allows to overcome thermalization bottlenecks and slow dynamics that have been observed in systems cooled by the coupling to boson baths.  
\end{abstract}

\maketitle


\section{Introduction}

Understanding the nonequilibrium properties of correlated lattice systems is relevant in connection with pump-probe experiments on solids and lattice modulation experiments realized in cold-atom setups. These experiments have revealed interesting phenomena, such as the switching to long-lived hidden states,\cite{Stojchevska2014} the realization of driving-induced topological phases,\cite{Jotzu2014} light-enhanced excitonic order,\cite{Mor2017} and -- perhaps most remarkably -- apparent light-induced superconducting states in various compounds.\cite{Kaiser2014,Hu2014,Mitrano2016} On the theory side, these developments have triggered efforts to interpret the measured phenomena,\cite{Denny2015,Okamoto2016,Kennes2017,Babadi2017,Murakami2017sc,Tanaka2018,Nava2018} to understand the mechanisms which govern the time-evolution of photo-excited systems,\cite{Eckstein2011,Golez2012,Sentef2013,Lenarcic2013,Golez2014,Murakami2015, Eckstein2016,Murakami2017ei,Golez2017,Ido2017} and to predict interesting nonequilibrium effects.\cite{Oka2009,Lindner2011,Mentink2015,Werner2015,Ono2017,Ono2018,Li2018} One difficulty encountered in such studies is that a strong light pulse injects energy into the system and typically results in heating. In systems with large gaps or internal degrees of freedom, some of this energy can be stored as potential energy for long times,\cite{Strand2017,Werner2018} but even there the heating is detrimental to the build-up of interesting quantum states, such as photo-induced Fermi liquids or photo-induced superconductors. 
Many theoretical works have thus considered lattice models coupled to fermionic or bosonic heat baths, 
\cite{Eckstein2013,Sentef2016,Murakami2017sc,Peronaci2019} 
in order to remove some of the energy injected by the pulse. This can be physically motivated, since photo-excited electrons in real solids interact with the spin background, phonons, and the electrons in other bands.\cite{DalConte2015,Rameau2016}  However, the (typically weak) coupling to the bath introduces new timescales which 
can make it difficult to reach the sought-after quantum states in simulations based on explicit time propagation. Moreover, in strongly correlated electron liquids, relaxation bottlenecks can appear which are intrinsic and apparently unrelated to the system-bath coupling.\cite{Eckstein2013,Sayyad2016} Such bottlenecks may invalidate protocols based on ``cooling after photodoping.'' 

Recently, it was shown that the photo-doping from narrow full bands, or into narrow empty bands, can produce a substantial cooling effect.\cite{Werner2019} This cooling results from a reshuffling of entropy from the lattice model into these narrow bands. To what extent such an effect may play a role in the experimentally reported photo-induced enhancements of electronic orders is an interesting open question. Here, we will not address this question, but rather exploit the fact that cooling-by-doping provides a powerful tool for inducing nonequilibrium states in lattice models which are relatively ``cold" -- without resorting to cold baths. Specifically, we will consider the paramagnetic single-band Hubbard model and revisit several questions which have been raised in previous studies: (i) Is there a difference in the optical conductivity of a chemically doped and photo-doped Mott insulator? (ii) Is it possible to induce a conventional $s$-wave superconducting state in the repulsive Hubbard model by producing a nonequilibrium state with a negative-temperature distribution? and (iii) Can an $\eta$-paired state be realized in a large-gap Mott insulator with a sufficiently high density of doublons and holons?

The paper is organized as follows. In Sec.~\ref{sec:method} we explain the method and the particle-hole symmetric set-up with dipolar coupling to full and empty narrow bands. In Sec.~\ref{sec:cond} we compare the conductivity of photo-doped Mott insulators to that of chemically doped equilibrium systems, in Sec.~\ref{sec:sc} we demonstrate the emergence of $s$-wave superconductivity in a metallic Hubbard model with a negative temperature distribution, and in Sec.~\ref{sec:eta} we show that the cooling-by-doping scheme allows to induce the $\eta$-paired state in a Mott system with a high density of cold doublons and holons. A summary and conclusions are given in Sec.~\ref{sec:conclusions}

\section{Method}
\label{sec:method}

We use the nonequilibrium extension of dynamical mean field theory (DMFT) \cite{Aoki2014} to simulate the nonequilibrium properties of the single-band Hubbard model with Hamiltonian
\begin{align}
H_\text{system}(t)=&\tilde v_\text{system}(t)\sum_{\langle i,j\rangle \sigma} (c^\dagger_{i\sigma} c_{j\sigma} + \text{h.c.})\nonumber\\
&+U\sum_{i}n_{i\uparrow}n_{i\downarrow}-\mu\sum_i(n_{i\uparrow}+n_{i\downarrow}),
\end{align}
where $c^\dagger_{i\sigma}$ creates a fermion on site $i$ with spin $\sigma$, $U$ is the on-site interaction, $\mu$ is the chemical potential, and $\tilde v$ the nearest-neighbor hopping. In DMFT this lattice model is mapped onto a quantum impurity model with interaction $U$ on the impurity and a bath characterized by the hybridization function $\Delta$. The latter is determined self-consistently in such a way that the bath mimics the lattice environment. We consider an infinitely connected Bethe lattice, where the self-consistency equation simplifies to\cite{Georges1996} 
\begin{equation}
\Delta_\sigma(t,t')=v_\text{system}(t)G_{\text{system},\sigma}(t,t')v^*_\text{system}(t).
\end{equation}
Here $G_\text{system}$ is the impurity Green's function and $v_\text{system}$ is a properly renormalized hopping amplitude.\cite{Metzner1989} For the solution of the impurity model, we use the non-crossing approximation (NCA).\cite{Keiter1971,Eckstein2010}

\begin{figure}[t]
\begin{center}
\includegraphics[angle=-90, width=0.8\columnwidth]{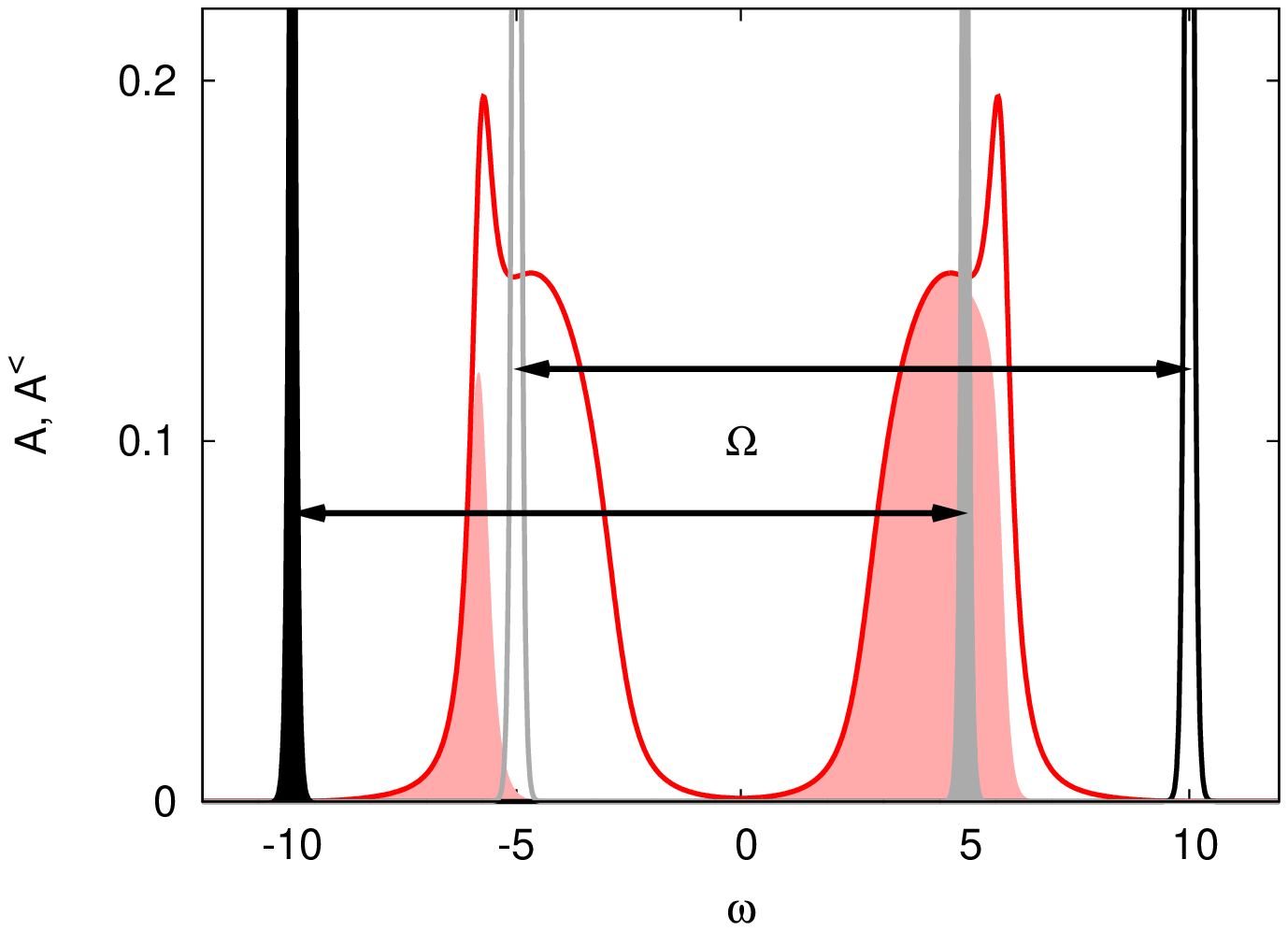}
\caption{Illustration of the symmetric photo-doping set-up with two narrow bands. The red (black) curves show the spectral function of the system (narrow bands). The upper Hubbard band is filled by injecting particles from the filled narrow band, with a driving frequency $\Omega$ appropriate for producing doublons. Simultaneously, the lower Hubbard band is emptied by ejecting particles into the empty narrow band, since the same driving energy couples the empty narrow band to the singly occupied states.  
}
\label{fig_illustration}
\end{center}
\end{figure}

In order to prepare cold nonequilibrium states, we transiently couple the system to two ($\alpha=$full,empty) narrow noninteracting bands, represented by corresponding noninteracting Green's functions $G^0_{\text{bath},\alpha}$. The hopping $v_\text{system-bath}(t)$ between system and bath is modulated in time with frequency $\Omega(t)$, 
\begin{equation}
v_\text{system-bath}(t)=a_\text{max}f_\text{envelope}(t)\sin\!\big(\Omega(t)t\big),
\end{equation}
with $a_\text{max}$ the maximum hopping amplitude and $f_\text{envelope}(t)$ an envelope function controlling the switch-on and switch-off of the pulse. This hopping modulation mimics dipolar excitations between the system and bath states.  
$\Omega(t)$ is adjusted in such a way that electrons are transferred from the full band to the system, and from the system to the empty band. Our symmetric set-up with two narrow bands is illustrated in Fig.~\ref{fig_illustration}. 

In the presence of the narrow bands, the self-consistency equation becomes
\begin{align}
&\Delta_{\text{system},\sigma}(t,t')=v_\text{system}(t)G_{\text{system},\sigma}(t,t')v^*_\text{system}(t)\nonumber\\
&\hspace{5mm}+\sum_\alpha v_\text{system-bath}(t)G^0_{\text{bath},\alpha}(t,t')v_\text{system-bath}^*(t').
\end{align}
The coupling to the narrow bands is switched on during a time controlled by $f_\text{envelope}$ and produces both the desired filling or occupation of the Hubbard bands, and a cooling effect. While the noninteracting narrow bands have the same temperature as the initial equilibrium system, it is the bandwidth rather than the temperature, which is relevant for the cooling.\cite{Werner2019} In particular, the effective temperature of the nonequilibrium state realized after the decoupling of the narrow bands can be substantially lower than that of the initial state. 

The above setup corresponds to an open system in which the narrow bands always remain in equilibrium. It has been shown in Ref.~\onlinecite{Werner2019} that similar cooling effects can be realized in a closed set-up, which involves separate self-consistency equations for the narrow bands. Since we merely use the doping from narrow bands here as a tool for realizing nonequilibrium states with a desired energy distribution, we choose the simpler open setup.

\begin{figure*}[ht]
\begin{center}
\includegraphics[angle=-90, width=0.66\columnwidth]{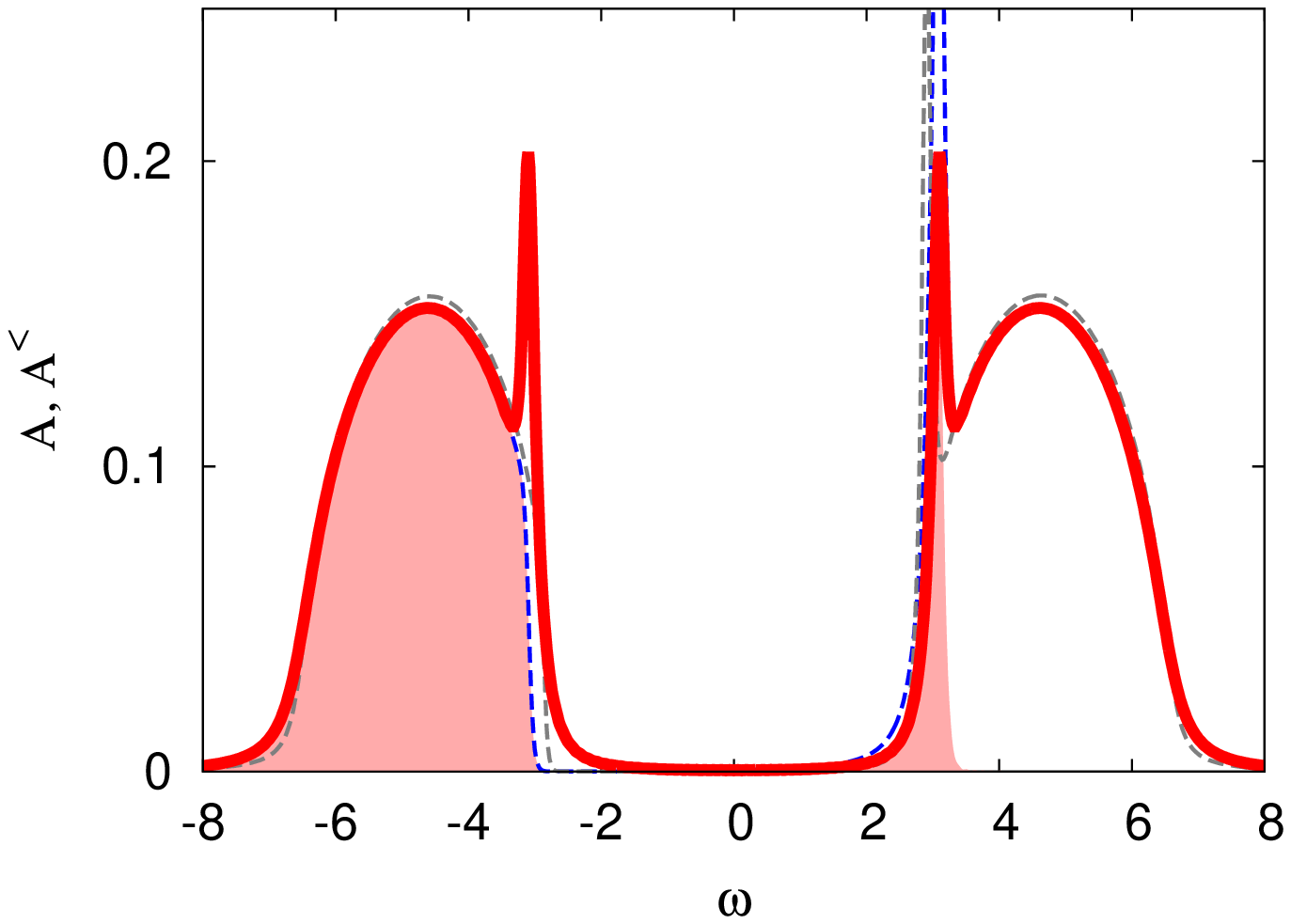}
\includegraphics[angle=-90, width=0.66\columnwidth]{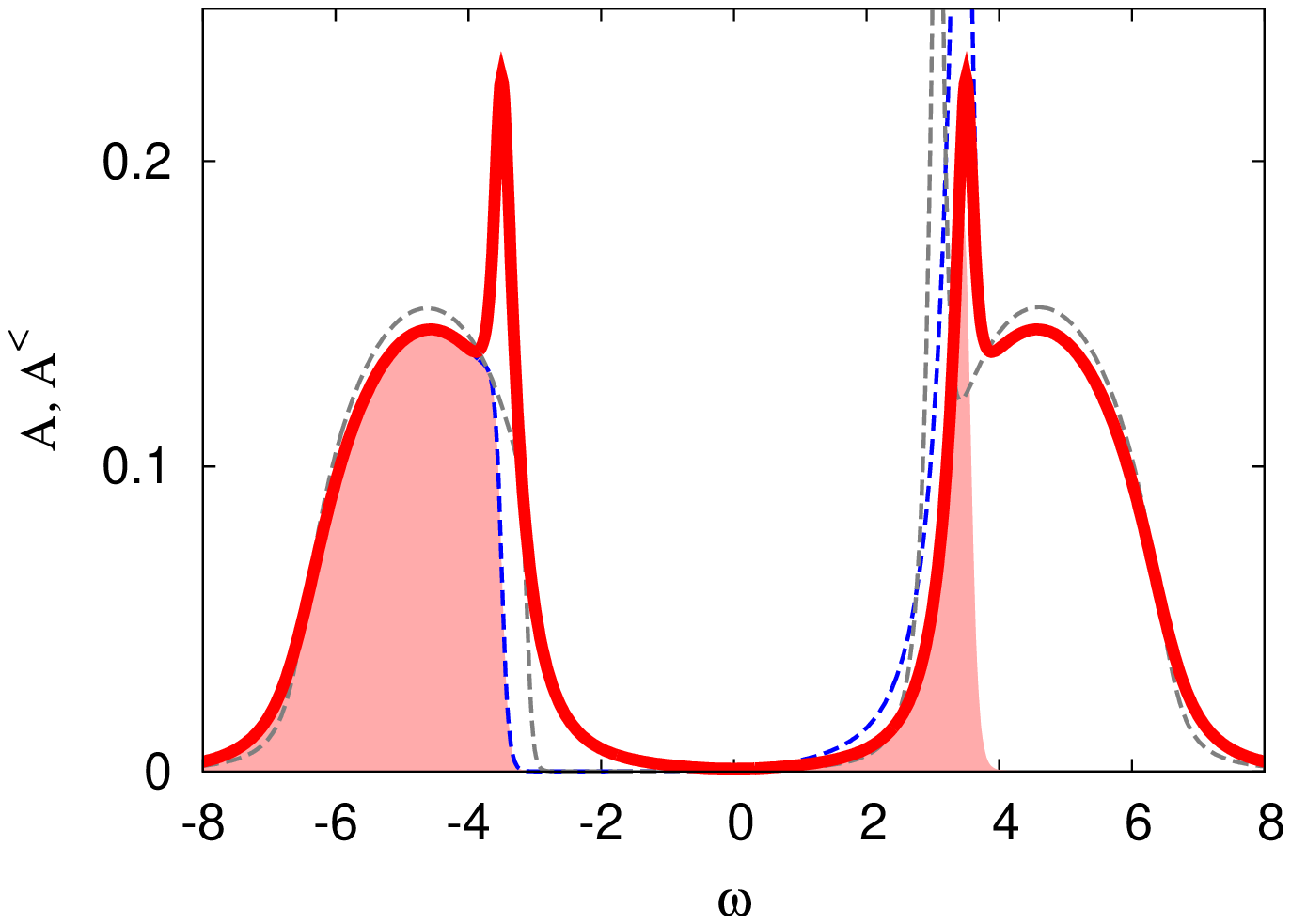}
\includegraphics[angle=-90, width=0.66\columnwidth]{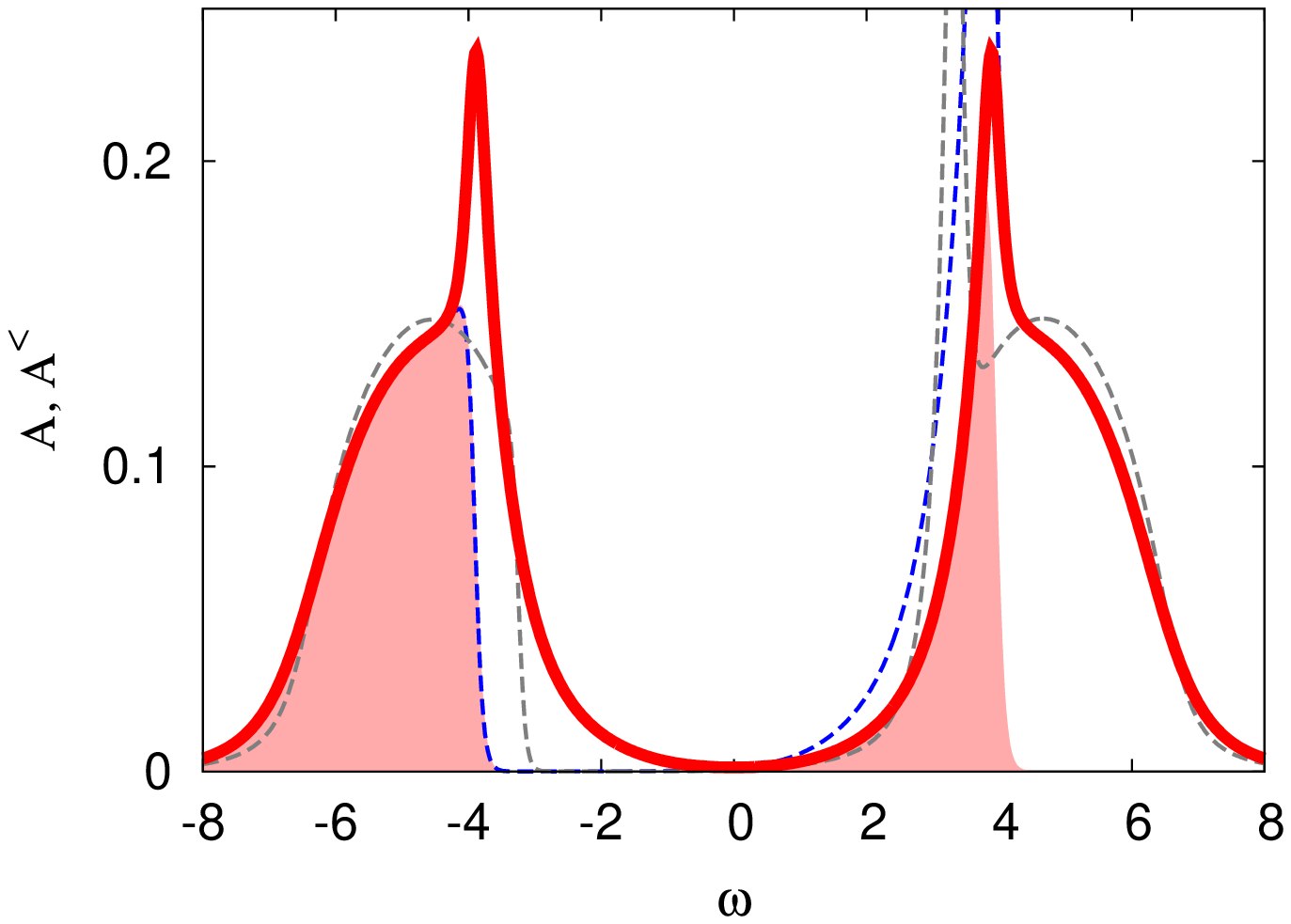}\\
\includegraphics[angle=-90, width=0.66\columnwidth]{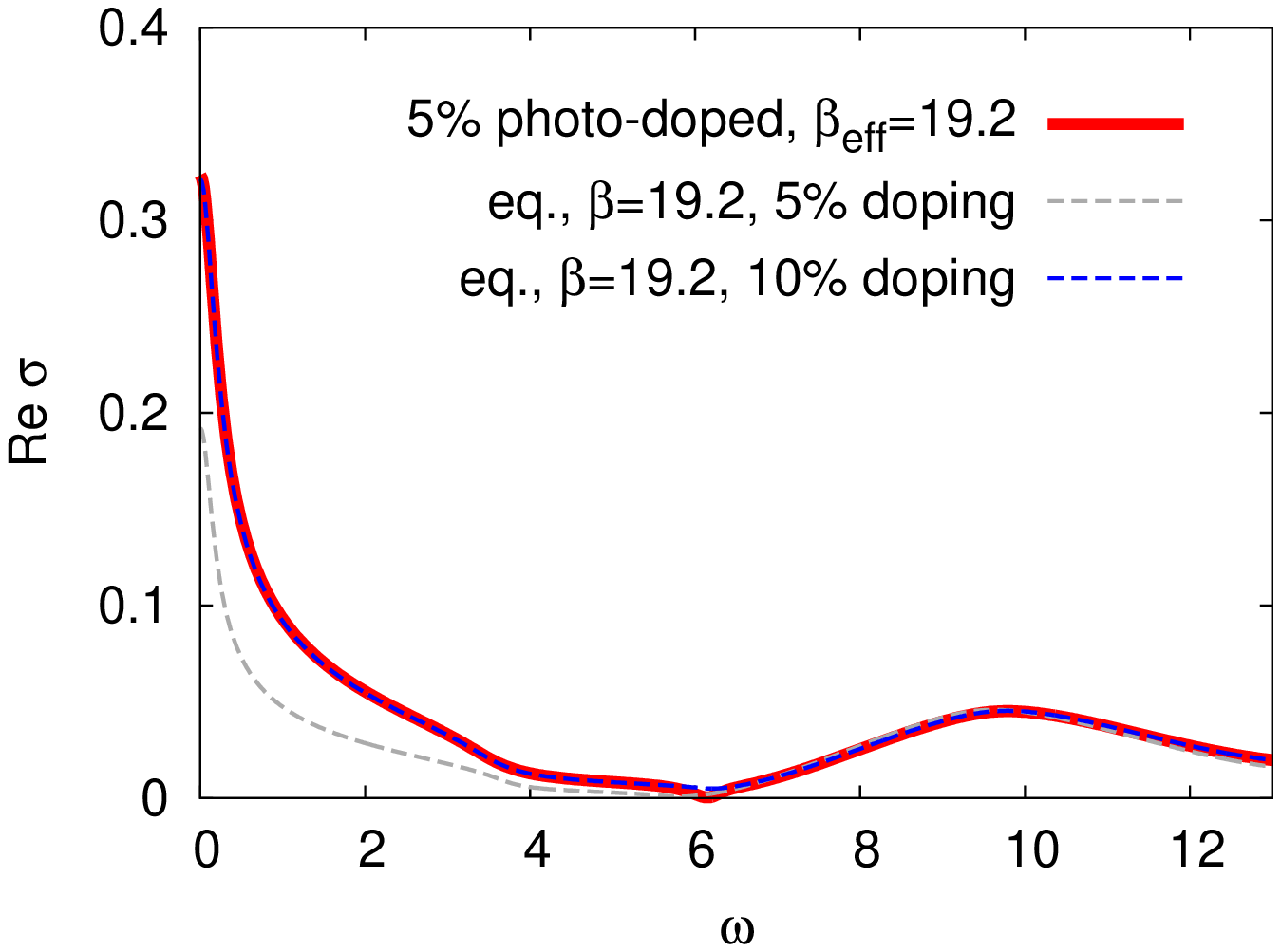}
\includegraphics[angle=-90, width=0.66\columnwidth]{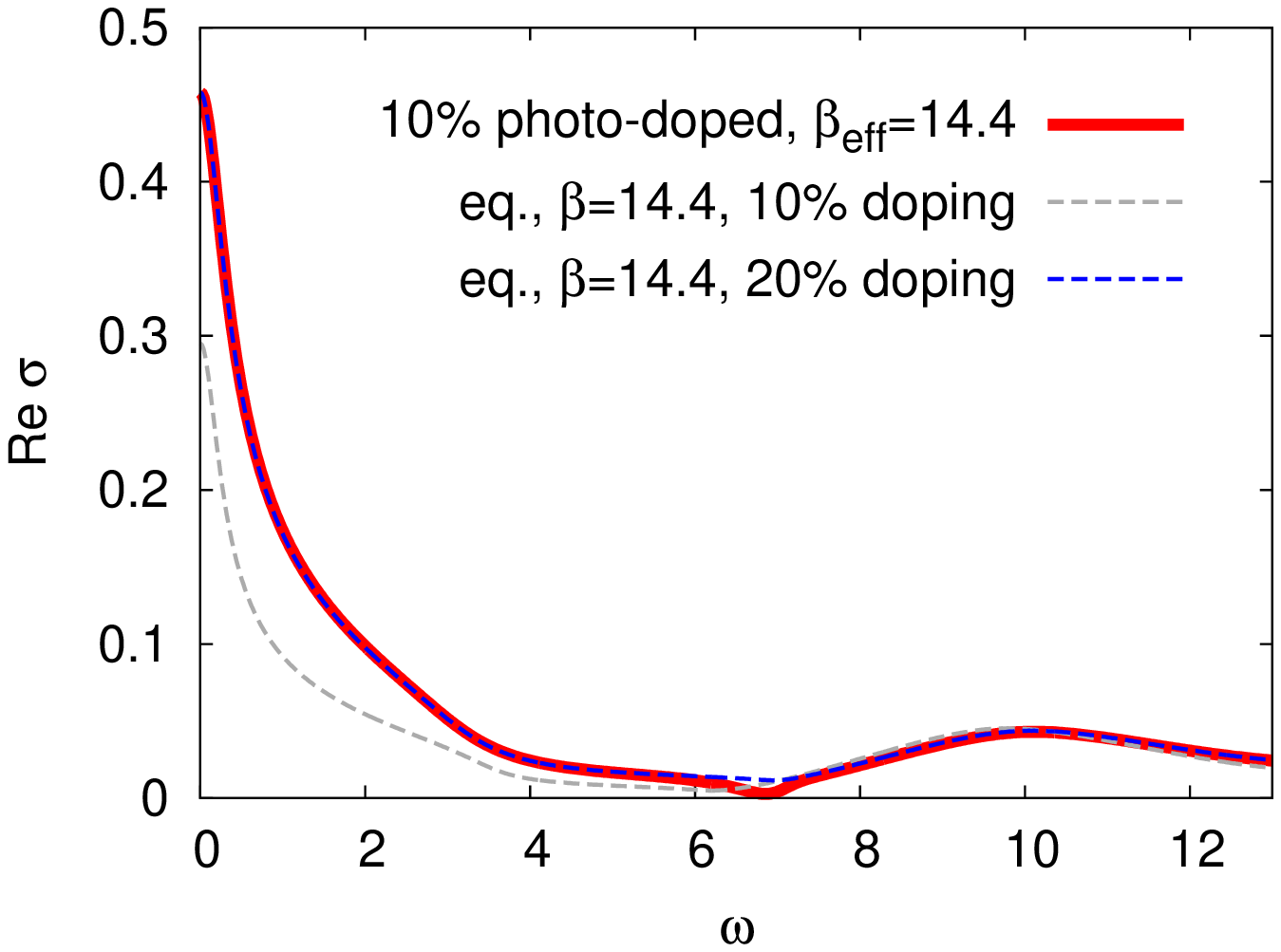}
\includegraphics[angle=-90, width=0.66\columnwidth]{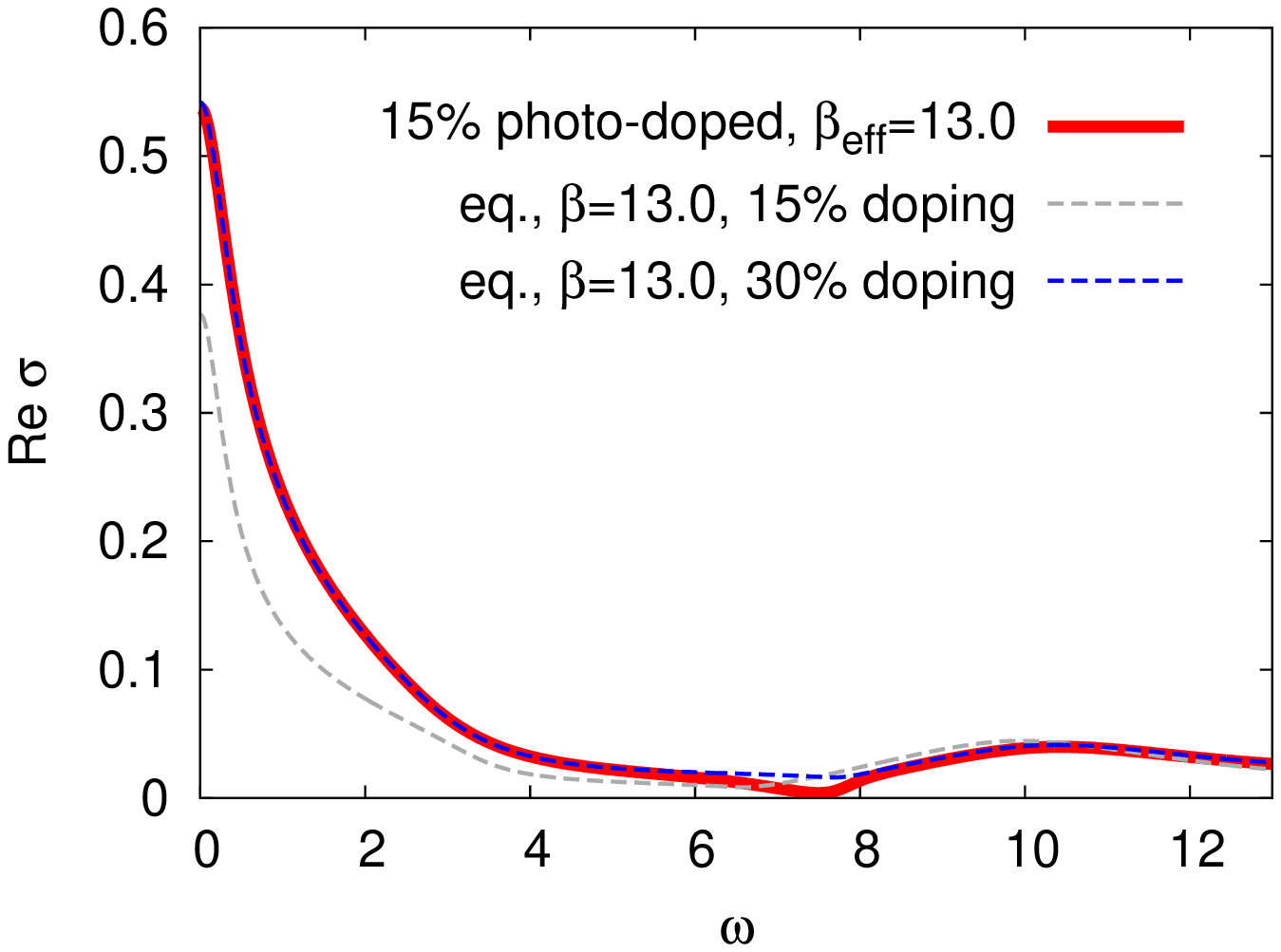}
\caption{Spectral functions (top) and optical conductivities (bottom) of photo-doped Mott insulators ($U=9$) with 5\%, 10\% and 15\% doublons and holons (from left to right). The grey (blue) dashed lines show equilibrium results for the effective temperature and chemical doping equal to the photo-doping (twice the photo-doping). For a better comparison, we have shifted the equilibrium spectra such that the positions of the Hubbard bands match those of the nonequilibrium spectra.  
}
\label{fig2}
\end{center}
\end{figure*}

To measure superconducting states, we introduce the Nambu formalism and the Green's function and hybridization function matrices
\begin{align}
&{\bf G}(t,t')=\left(
\begin{array}{ll}
G^{cc^\dagger}_{\uparrow \uparrow}(t,t') & G^{cc}_{\uparrow \downarrow}(t,t')\\
G^{c^\dagger c^\dagger}_{\downarrow \uparrow}(t,t') & G^{c^\dagger c}_{\downarrow \downarrow}(t,t')
\end{array}
\right),\\
&{\bf \Delta}(t,t')=\left(
\begin{array}{ll}
\Delta^{c^\dagger c}_{\uparrow \uparrow}(t,t') & \Delta^{c^\dagger c^\dagger}_{\uparrow \downarrow}(t,t')\\
\Delta^{c c}_{\downarrow \uparrow}(t,t') & \Delta^{c c^\dagger}_{\downarrow \downarrow}(t,t')
\end{array}
\right).
\end{align}
With these matrices, and the Pauli matrix ${\bf \sigma}_3$, the self-consistency equation (without bath) for conventional $s$-wave superconductivity can be written as 
\begin{equation}
{\bf \Delta}_\text{sc}(t,t')=v_\text{system}(t){\bf \sigma}_3 {\bf G}(t,t') {\bf \sigma}_3 v^*_\text{system}(t').\label{sc}
\end{equation}
$\eta$-pairing corresponds to a staggered form of $s$-wave pairing,\cite{Yang1989} which on the bipartite Bethe lattice can be realized by changing the sign of the off-diagonal (anomalous) components in the self-consistency equation, which becomes
\begin{equation}
{\bf \Delta}_\eta(t,t')=v_\text{system}(t) {\bf G}(t,t') v^*_\text{system}(t').\label{eta}
\end{equation}

After decoupling from the the narrow bands, the $\eta$-pairing order parameter is conserved under the time-evolution. To enable a nontrivial dynamics, one can add a next-nearest neighbor hopping. An approximate treatment of the self-consistency for the bipartite Bethe lattice then yields the equation\cite{Georges1996}
\begin{align}
&{\bf \Delta}_{\eta,\text{NNN}}(t,t')=v_\text{system}^\text{NN}(t) {\bf G}(t,t') (v_\text{system}^\text{NN}(t'))^*\nonumber\\
&\hspace{10mm}+v_\text{system}^\text{NNN}(t){\bf \sigma}_3 {\bf G}(t,t') {\bf \sigma}_3 (v_\text{system}^\text{NNN}(t'))^*,
\end{align}
with $v_\text{system}^\text{NN}$ ($v_\text{system}^\text{NNN}$) corresponding to the nearest-neighbor (next-nearest-neighbor) hopping, respectively. 

To evaluate the optical conductivity $\sigma$ (which is done for simplicity only for the system with $v_\text{system}^\text{NNN}=0$), we define ${\bf v}_\text{system}$ in the selfconsistency equations (\ref{sc}) and (\ref{eta}) as a diagonal hopping matrix whose components are multiplied with appropriate Peierls factors:\cite{Aoki2014,Werner2017} ${\bf v}_\text{system}(t)=\text{diag}(v_\text{system}e^{i\phi(t)}, v_\text{system}e^{-i\phi(t)})$ with  $\phi(t)=-\int_0^t ds E(s)ea/\hbar c$ corresponding to an electric field pulse $E(t)$ ($a$ is the lattice spacing, $e$ the electron charge, and $c$ the speed of light, all of which are set to one in the following). The selfconsistency equations (\ref{sc}) and (\ref{eta}) are then modified as ${\bf \Delta}_\text{sc}(t,t')$ $=$ $\frac{1}{2}[{\bf v}_\text{system}(t){\bf \sigma}_3 {\bf G}(t,t') {\bf \sigma}_3 {\bf v}^*_\text{system}(t')$ $+$ ${\bf v}_\text{system}^*(t){\bf \sigma}_3 {\bf G}(t,t') {\bf \sigma}_3 {\bf v}_\text{system}(t')]$ and ${\bf \Delta}_\eta(t,t')$ $=$ $\frac{1}{2}[{\bf v}_\text{system}(t){\bf G}(t,t'){\bf v}^*_\text{system}(t')$ $+$ ${\bf v}_\text{system}^*(t){\bf G}(t,t'){\bf v}_\text{system}(t')]$. 

The induced current $j(t)$ can be measured as follows:
\begin{equation}
j(t) = \text{Im}\text{Tr}[\sigma_3{\bf\Gamma}_\text{L-R}(t)],
\end{equation}
with ${\bf \Gamma}_\text{L-R}(t)=-i[{\bf G}\ast{\bf \Delta}_\text{L-R}]^<(t,t)$ and ${\bf \Delta}_\text{L-R}$ given by  ${\bf \Delta}_\text{sc,L-R}(t,t')$ $=$ $\frac{1}{2}[{\bf v}_\text{system}(t){\bf \sigma}_3 {\bf G}(t,t') {\bf \sigma}_3 {\bf v}^*_\text{system}(t')$ $-$ ${\bf v}_\text{system}^*(t){\bf \sigma}_3 {\bf G}(t,t') {\bf \sigma}_3 {\bf v}_\text{system}(t')]$ 
and ${\bf \Delta}_{\eta,\text{L-R}}(t,t')$ $=$ $\frac{1}{2}[{\bf v}_\text{system}(t){\bf G}(t,t'){\bf v}^*_\text{system}(t')$ $-$ ${\bf v}_\text{system}^*(t){\bf G}(t,t'){\bf v}_\text{system}(t')]$, respectively. Finally, to obtain the time- and frequency dependent conductivity, we choose a half-cycle electric field pulse $E_{t'}(t)$ centered at a time $t=t'$. From the Fourier transformations of $E_{t'}(t)$ and the resulting current $j_{t'}(t)$, one obtains $\sigma(\omega,t')=j_{t'}(\omega)/E_{t'}(\omega)$. Here, we perform a forward-in-time Fourier transformation starting at at time smaller than $t'$, and in the case of the current use a Fermi-function-like cutoff at the longest simulation times. 
In the following calculations, we will use $v_\text{system}$ $(\hbar/v_\text{system})$ as the unit of energy (time). Without coupling to the narrow bands, the system thus has a noninteracting density of states which is semi-circular with a bandwidth of $4$.

The time-dependent spectral function and occupation function are calculated from the retarded and lesser Green's functions via forward Fourier transformation:
\begin{align}
A(\omega,t)&=-\frac{1}{\pi}\text{Im}\int_t^{t_\text{max}} dt'e^{i\omega (t'-t)}G^\text{ret}(t',t),\\
A^<(\omega,t)&=\frac{1}{\pi}\text{Im}\int_t^{t_\text{max}} dt'e^{i\omega (t'-t)}G^<(t',t).
\end{align}

\section{Results}

\subsection{Optical conductivity of photo-doped states}
\label{sec:cond}

Photodoping a Mott insulator should produce a metallic phase, but, remarkably, in explicit time dependent simulations this phase turned out to be a bad metal rather than a Fermi liquid. Although the precise mechanism is yet unclear, in the previous DMFT simulations it seems to be related to a combination of strong heating upon photodoping, and an inefficient process of cooling the resulting state down to the Fermi liquid regime.  Ref.~\onlinecite{Eckstein2013} showed that photo-doped Mott insulators, even if cooled by a phonon bath, do not exhibit Fermi liquid behavior and sharp quasi-particle peaks up to the longest numerically accessible times. While the phonon bath is efficient in reducing the kinetic energy of the system, there appears to be a bottleneck in the formation of the quasi-particles. As a result, the nonequilibrium spectral functions of the weakly photo-doped system showed no clear signature of a quasi-particle band, and the Drude feature in the optical conductivity remained very weak and broad. A possibly related bottleneck was observed in the vicinity of the Mott phase after hopping quenches.\cite{Sayyad2016} 

The strong heating after direct photo-excitation of electrons across the Mott gap can be understood as a consequence of the large entropy of the paramagnetic Mott insulator in DMFT: the $\ln(2)$ entropy per site in the initial Mott insulating state is larger than in a low-temperature Fermi liquid state of the chemically doped equilibrium system. The photo-doping will increase the entropy further and thus result (before cooling) in a very hot state with a broad energy distribution of the doublons and holons in the respective Hubbard bands.\cite{Eckstein2011} In view of these results, one may ask whether the photo-doped Mott insulator, in contrast to the chemically doped one, can be turned into a Fermi liquid {\em at all}. Ref.~\onlinecite{Werner2019}, which introduced the cooling-by-doping protocol, mentioned that the answer was affirmative. In the following, we provide a more in-depth analysis of this issue. 

We start from a half-filled Mott insulator with $U=9$ and initial inverse temperature $\beta=5$ and inject doublons into the upper Hubbard band (holons into the lower Hubbard band) from a filled (empty) narrow band by dipolar excitations (Fig.~\ref{fig_illustration}). The narrow bands have a box-shaped density of states of width $0.05$ with Fermi-function like edges corresponding to a ``cutoff temperature" $0.01$. The duration of the excitation pulse is $\Delta t\approx 50$ and the amplitude is $a_\text{max}=0.15$ (left panels) or 0.25 (middle and right panels). The effective temperature reached after the photo-doping depends on the details of the chirping protocol $\Omega(t)$, but we can easily prepare photo-doped states with effective temperature substantially below $\beta=5$ (the precise pulse forms used are listed in Appendix~\ref{app}). These effective temperatures, which are determined from the energy distributions of the doublons and holons in the respective Hubbard bands, are cold enough that sharp quasi-particle peaks emerge near the edges of the Mott gap. 

In Fig.~\ref{fig2} we plot results for 5\%, 10\%, and 15\% photo-doping. Here, $x$\% photo-doping means $x$\% photo-doped doublons and $x$\% photo-doped holons. The effective temperatures extracted from a Fermi-function fit to $A^</A$ in the energy range of the upper Hubbard band are $\beta_\text{eff}=19.2$, $14.4$, and $13$, respectively. The spectral function and conductivity of the nonequilibrium system, i.e. for the long-lived photo-doped state realized after the pulse, are plotted by thick red lines, with the red shading indicating the occupation ($A^<$). We also plot equilibrium results corresponding to $\beta=\beta_\text{eff}$ and $x$\% (gray) and $2x$\% (blue) chemical doping.  The Drude peak of the nonequilibrium conductivity matches within the accuracy of the calculation the equilibrium result for $2x$\% hole doping. In other words, a photo-doped state with $x\%$ doublons and $x\%$ holons has, within DMFT, the same low-frequency conductivity as a chemically doped system with $2x\%$ doping and $\beta=\beta_\text{eff}$. The only significant difference between the photo-doped and chemically doped system is a bleaching effect at frequencies $6\lesssim \omega\lesssim  7.5$, where excitations in the photo-doped system are suppressed because of the nonthermal population. This bleaching effect becomes more pronounced with stronger photo-doping, and for a large enough population inversion, the conductivity of the photo-doped state becomes negative.  

The spectral functions of the photo-doped and chemically doped system are of course different, since the latter exhibit only one quasi-particle band, at the edge of the upper Hubbard band (for electron doping), while the former has quasi-particle bands associated with both doublons and holons. In Fig.~\ref{fig2} we have shifted the spectral functions of the chemically doped systems in such a way that the Hubbard bands approximately match those of the photo-doped system. Interestingly, the shape of the lower Hubbard band in $A(\omega)$ for the chemically doped system with 2$x$\% doping almost perfectly matches the occupation $A^<(\omega)$ of the photo-doped system. This indicates an identical distribution of singly occupied states in both systems, further confirming the close relation between the photo-doped and chemically doped states.

\subsection{Superconductivity in negative-temperature states}
\label{sec:sc}

A remarkable prediction from a nonequilibrium DMFT study of the Hubbard model \cite{Tsuji2011} is that an AC field quench in the metallic phase renormalizes the effective Coulomb interaction as
\begin{equation}
U_\text{eff}=\frac{U}{J_0(E/\Omega)}, 
\end{equation}
with $\Omega$ the driving frequency, $E$ the field amplitude and $J_0$ the zero-order Bessel function. Hence, if $E/\Omega$ is chosen such that $J_0(E/\Omega)<0$, the effective interaction becomes attractive. This result is based on the fact that an AC field quench with such an $E/\Omega$ effectively flips the band and results in a population inversion in the flipped band. Since a repulsively interacting system with flipped band and a negative temperature distribution is equivalent to an attractively interacting system with a positive temperature distribution, one obtains the above formula for the effective interaction.\cite{Tsuji2011} The interesting question is if this mechanism allows to induce $s$-wave superconductivity in a repulsively interacting system. 

\begin{figure}[t]
\begin{center}
\includegraphics[angle=-90, width=0.9\columnwidth]{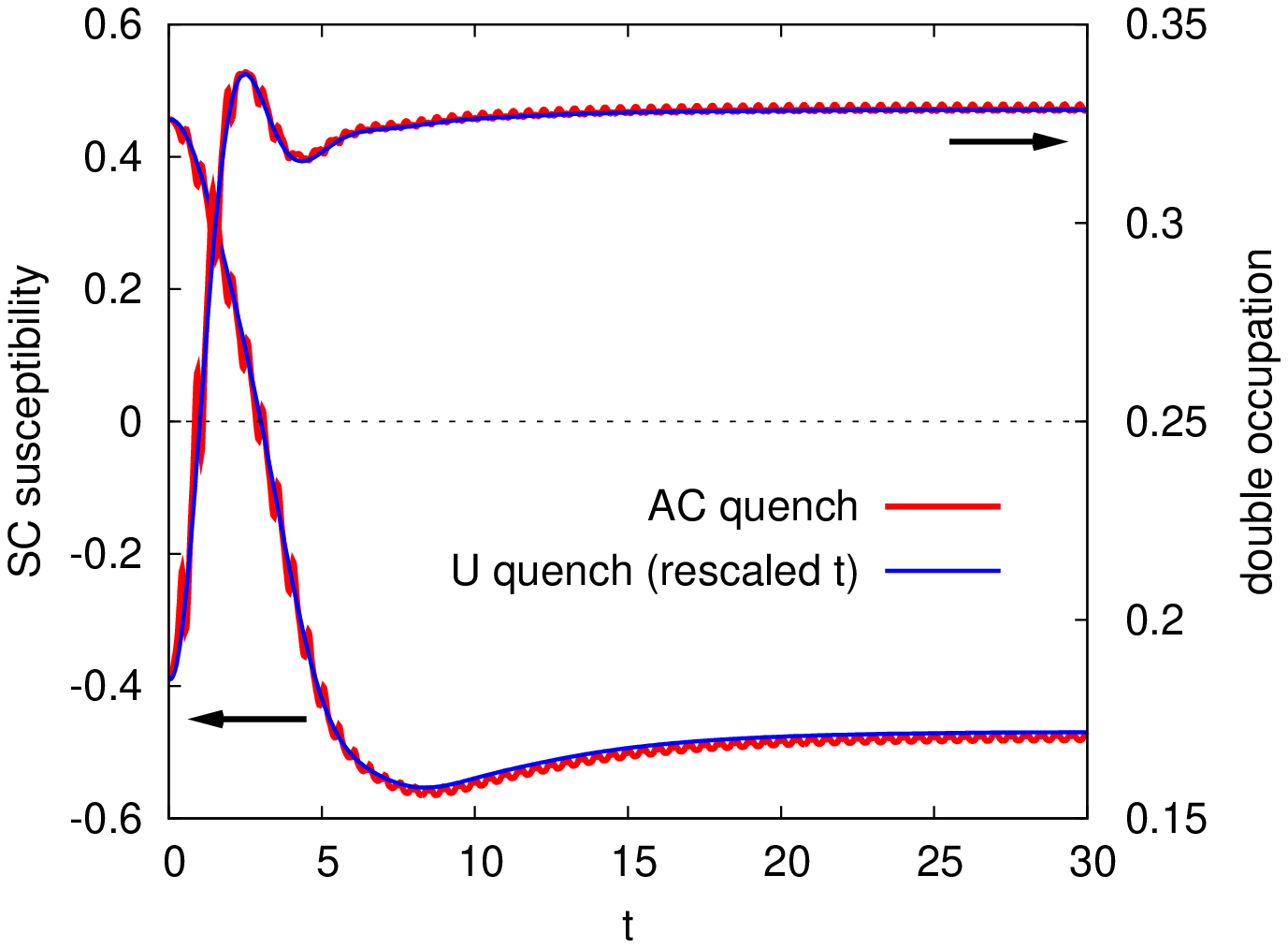}
\caption{Double occupation and pair susceptibility after an AC field quench with $\Omega=2\pi$, $E=4$ (red lines) and an equivalent interaction quench (blue lines). The parameters of the initial state are $U=1$, $\beta=5$. 
}
\label{fig_tsuji}
\end{center}
\end{figure}

\begin{figure}[b]
\begin{center}
\includegraphics[angle=-90, width=0.8\columnwidth]{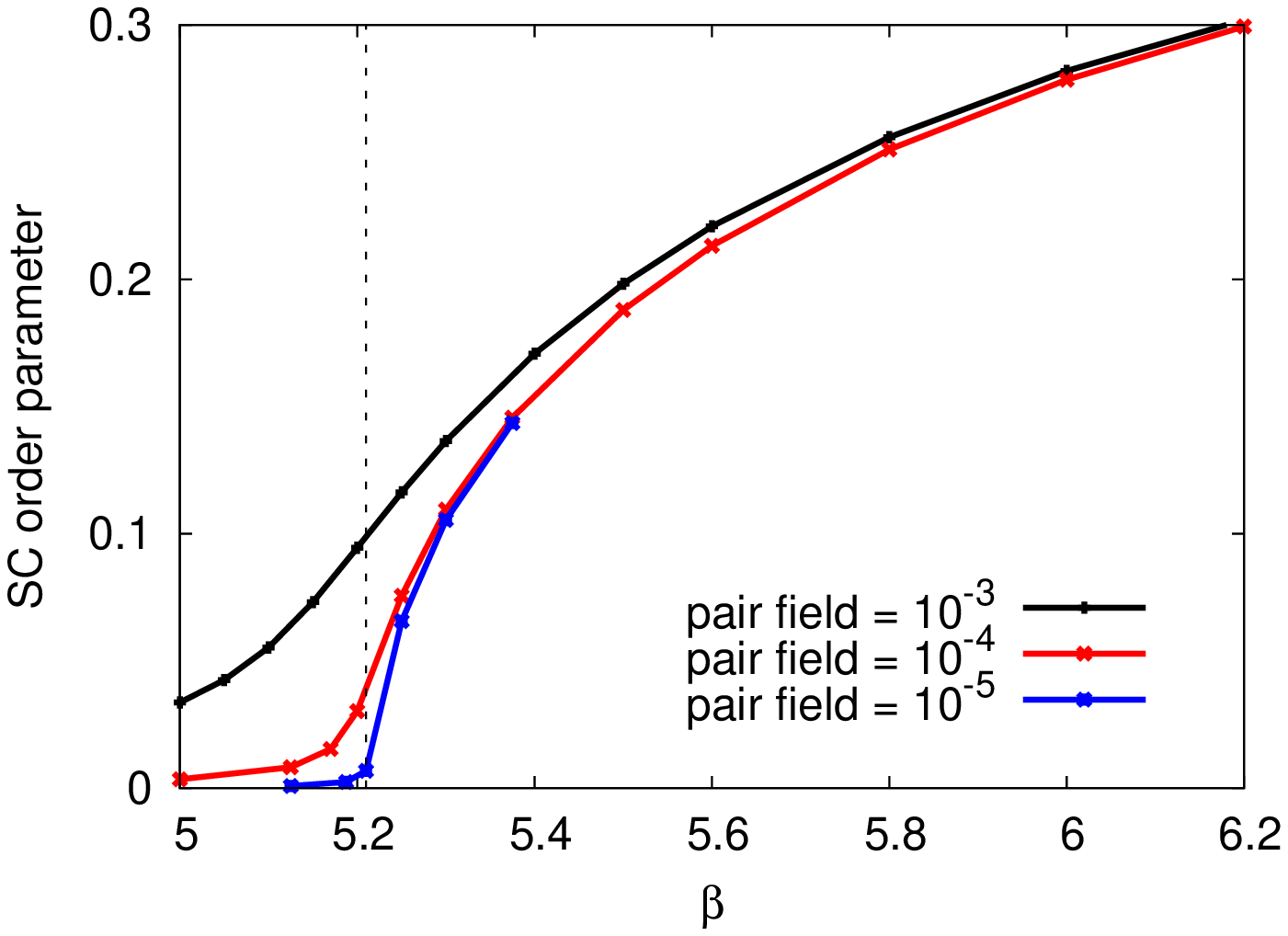}
\caption{Superconducting order parameter as a function of $\beta$ at $U=-2.52$ for indicated values of the applied pair field.}
\label{fig_seed}
\end{center}
\end{figure}

We can check the effect of the AC field quench on the pairing susceptibility by  applying a small pair field and measuring the induced order parameter. Figure~\ref{fig_tsuji} shows the results for $\Omega=2\pi$, $E=4$ and initial $U=1$, $\beta=5$. For these driving parameters, we expect a dynamics which effectively corresponds to a quench of the initially repulsive interaction to an effective interaction $U_\text{eff}=-2.52$. (Precisely speaking, the AC quench corresponds to a hopping quench to $v_\text{system}=-1/2.52$, so in the comparison to the $U$ quench, we need to rescale the time axis by $2.52$ and the pair field by $-2.52$.) The corresponding $U$-quench data are shown by the blue lines. Indeed, we see that the results for the $E$-field driving and the $U$-quench agree up to small oscillations. In the limit of $\Omega\rightarrow \infty$, the two results would perfectly match. 

\begin{figure*}[ht]
\begin{center}
\includegraphics[angle=-90, width=0.66\columnwidth]{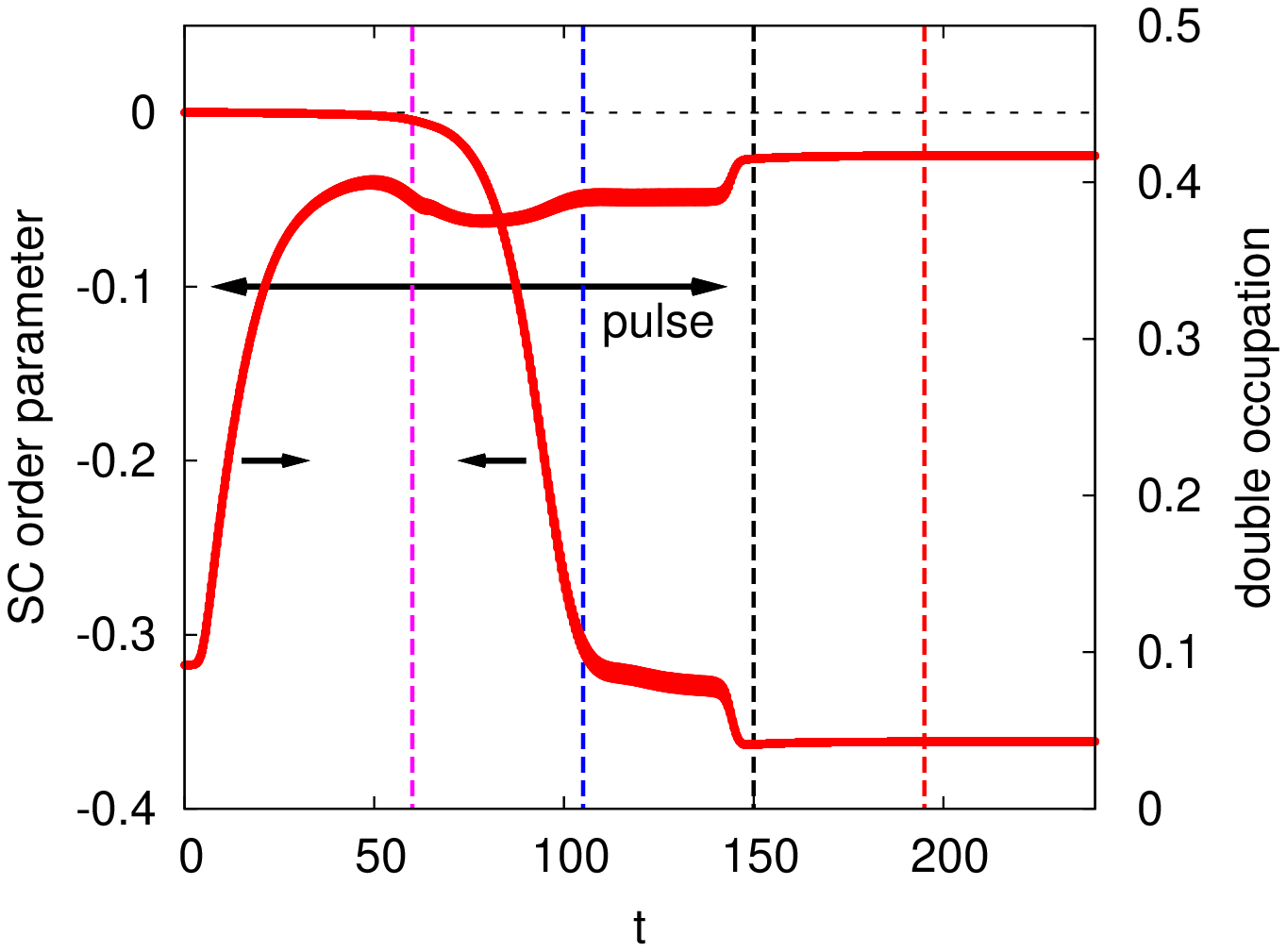}
\includegraphics[angle=-90, width=0.66\columnwidth]{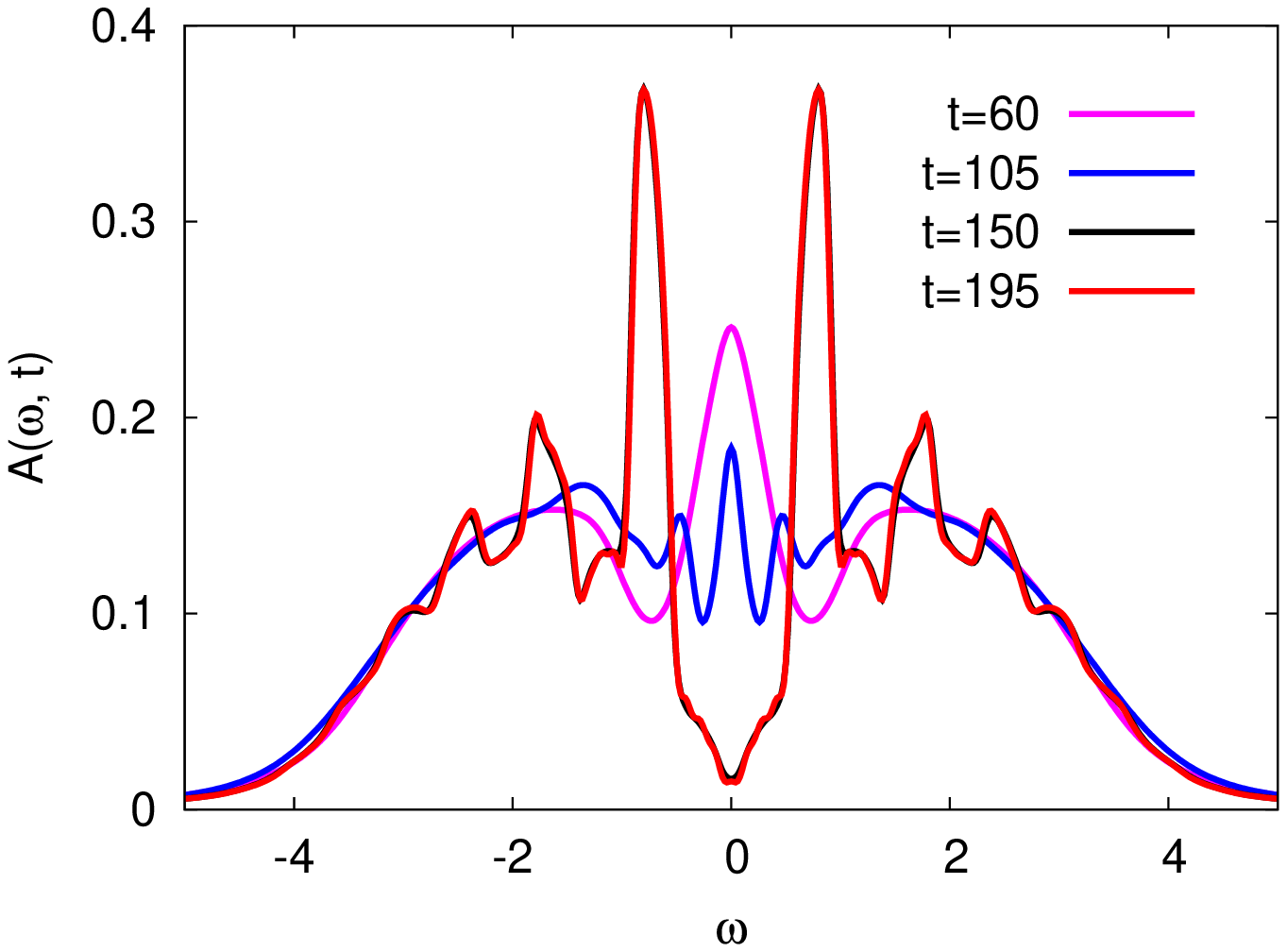}
\includegraphics[angle=-90, width=0.66\columnwidth]{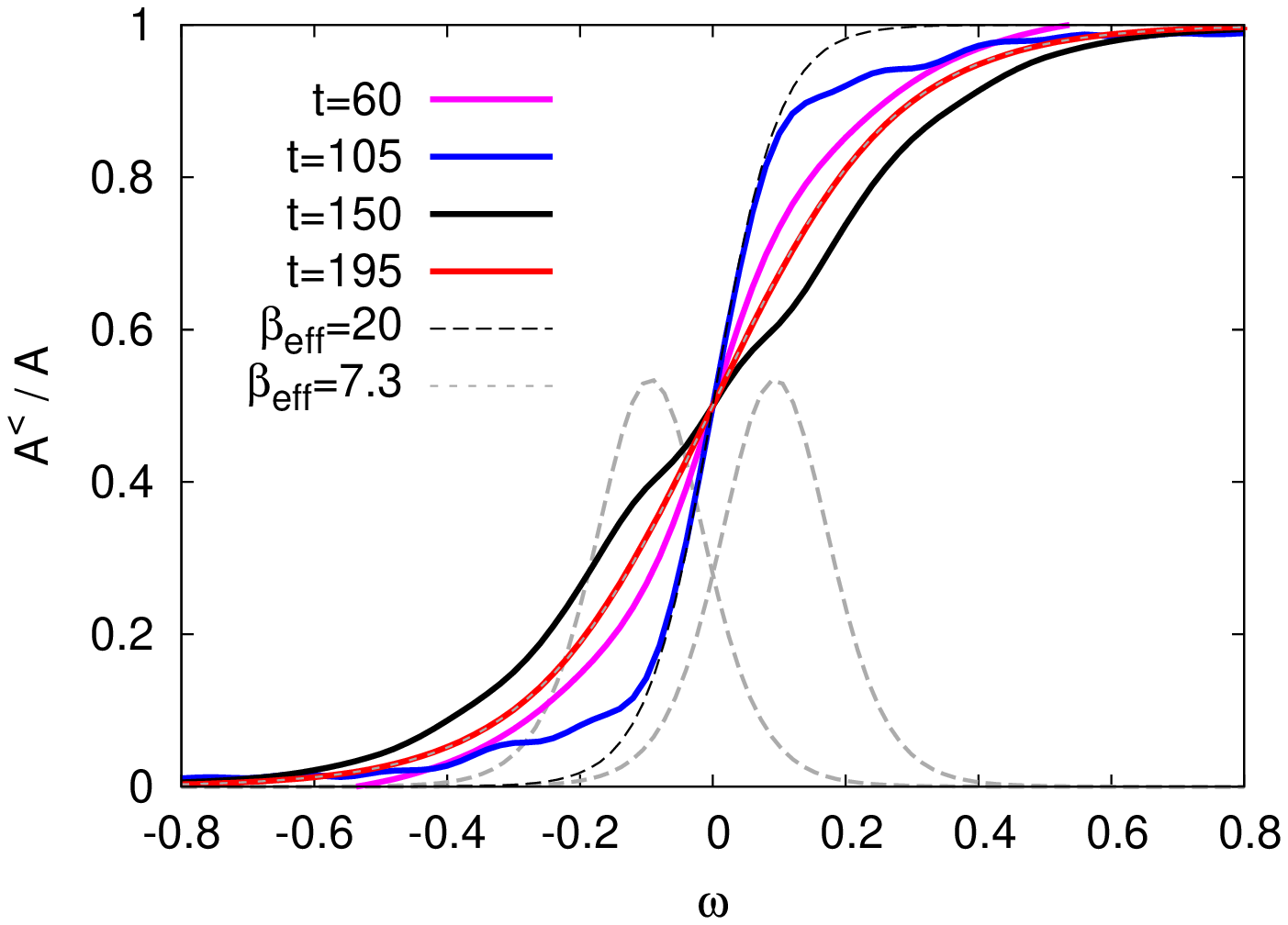}
\caption{Left panel: Double occupation and superconducting order parameter during the symmetric cooling-by-doping process. The interaction strength is $U=2.52$, the initial inverse temperature $\beta=5$, and the applied pair field is $0.001$. Middle and right panels: Spectral functions $A$ and distribution functions $A^</A$ measured at the times indicated by the vertical lines in the left panel. The dashed gray lines in the right panel represent the density of states of the flat bands shifted by the driving frequency for $t>94$.  
}
\label{fig_inverted}
\end{center}
\end{figure*}

We notice that the double occupation increases beyond $0.25$, which implies a population inversion, consistent with the results in Ref.~\onlinecite{Tsuji2011}, which were obtained using a numerically exact impurity solver.\cite{Werner2009} The susceptibility changes sign, because the mapping from negative to positive temperature changes the sign of the pair field.  
There is no strong enhancement in the susceptibility, even though $U_\text{eff}=-2.52$ with a critical $\beta_c\approx 5.21$ is close to the optimal $U$ in the NCA phase diagram.\cite{Werner2012} The reason for the low pair susceptibility is that $|\beta_\text{eff}|$ after the AC or $U$-quench is rather small ($\beta_\text{eff}=-0.57$), i.e. the system has a very ``hot" negative temperature distribution.  

As discussed in Ref.~\onlinecite{Tsuji2011}, $|T_\text{eff}|$ can be substantially reduced ($|\beta_\text{eff}|$ enhanced) by implementing a multi-step quench. For example, in the simulation with the NCA solver, if we quench from $U=0.1$ to $-0.252$ at $t=0_+$, hold the interaction at $-0.252$ up to $t=6$, and then ramp it linearly to $-2.52$ in a time $t_\text{ramp}=60$, the inverse temperature of the thermalized system with $U=-2.52$ is $\beta=3.9$. With further optimizations of the multi-step quench protocol it may be possible to realize the superconducting state. 

Here, we want to explore if cooling-by-photo-doping allows us to produce a negative-temperature state 
in the repulsive Hubbard model with $U=2.52$ which is sufficiently cold that the system becomes superconducting. For this study it is important to use the symmetric doping scheme sketched in Fig.~\ref{fig_illustration} because it avoids a mismatch in chemical potential between the original and photo-doped system, which would be detrimental to superconductivity. In the symmetric scheme, the population inversion is realized by promoting singly occupied sites to doubly occupied sites with high kinetic energy (insertion of particles from the full narrow band), and at the same time removing singly occupied sites with low kinetic energy (ejection of particles into the narrow empty band). It is also useful for the interpretation of the following data to have a look at the equilibrium temperature dependence of the superconducting order parameter for $U=-2.52$ and different weak pair fields (Fig.~\ref{fig_seed}). At this interaction strength, the symmetry breaking occurs near $\beta_c\approx 5.21$. With an applied pair field of 0.001, however, we need to realize an order parameter $>0.1$ to claim that the system exhibits a spontaneous symmetry breaking. 

Figure~\ref{fig_inverted} shows the results for a cooling-by-photodoping simulation, in which the frequency of the pulse is continuously lowered. The idea is to initially add particles at the upper edge of the spectral function (remove particles at the lower edge), and then, as the population inversion is building up, slowly shift the narrow bands closer to $\omega=0$.  
Here we choose narrow bands with a box-like density of states of width $0.1$ and Fermi-function like edges with cutoff temperature $0.05$. The left panel shows the time evolution of the double occupation $d$ and the superconducting order parameter induced by a pair field of strength 0.001. The double occupation shows that the pulse excitation quickly realizes an inverted population with $d>0.25$. The order parameter in the driven system exhibits a roughly exponential growth once the population inversion is realized. It quickly grows in magnitude beyond 0.1 (the order parameter is negative, because the inverted population implies that the sign of the pair field is effectively inverted). The superconducting order parameter saturates with a magnitude of about 0.33 in the driven state with coupling to the narrow bands, and increases to about 0.36 after the decoupling from these bands, and the thermalization of the system in the negative-temperature state.  

In the absence of dissipation, the negative-temperature state is stable, and hence we have demonstrated the possibility of inducing conventional $s$-wave superconductivity in such a nonequilibrium state in a repulsive Hubbard model. 
This superconducting state is equivalent to the equilibrium $s$-wave superconducting state in the attractive Hubbard model.\cite{Tsuji2011} It is interesting to also consider the spectral functions and energy distribution functions, which are plotted in the middle and right panel for the times marked by the vertical dashed lines in the left panel. Even though the distribution function is not exactly Fermi-like, the steepest slope of the distribution function at $\omega=0$ (corresponding to $\beta_\text{eff}=20$) is reached in the saturated state with coupling to the narrow bands (blue line). Here the pulse frequency is such that particles are inserted at an energy slightly above $\omega=0$, while particles are removed at energies slightly below $\omega=0$, see dashed gray lines for the positions and widths of the shifted narrow bands. The blue distribution function thus corresponds to a nonequilibrium ``steady state" characterized by a flow of particles from the full band via the system to the empty band, and the steep distribution function is a consequence of the two narrow bands being shifted by the drive into close proximity of the Fermi level. While the state is superconducting, the spectral function of the steady state does not exhibit a gap. This is a result of the strong hybridization to the (normal) baths at low energy, i.e. in the gap region. 

After the system is decoupled from the narrow bands, a thermalization process in the population-inverted state sets in, resulting in an inverse temperature of $\beta=-7.3$. This is cold enough for superconductvity, and indeed, the superconducting gap in the spectral function opens almost immediately after the decoupling, see black and red spectra. The shape of the spectral function is consistent with the equilibrium result for $\beta=7.3$ and an attractive interaction of $U=-2.52$.

\subsection{$\eta$-pairing in strongly photo-doped Mott states}
\label{sec:eta}

The negative temperature superconducting state should also exist in the Mott regime, but the population inversion within the upper and lower bands is expected to make it unstable against coupling to equilibrium heat baths. There is however another type of superconducting state, which can be realized in Mott insulators with a large density of doublons and holons and a positive effective temperature, the so-called $\eta$-paired state.\cite{Yang1989,Rosch2008} This superconducting state is characterized by a staggered order parameter, with a sign change between the two sublattices. 

The life-time of doublons and holons grows exponentially with the gap size,\cite{Sensarma2010,Eckstein2011,Rosch2008}  
and for a sufficiently large gap, we can neglect heating from doublon-holon recombination processes. Hence, if it is possible to produce a nonequilibrium state with a large density of doublons and holons and a cold enough effective temperature, a symmetry breaking to the $\eta$-paired state may be induced. A recent time-dependent exact diagonalization study of a small system \cite{Kaneko2019} has demonstrated a strong photo-induced enhancement of $\eta$-pairing correlations. An interesting open question is if similar photo-doping pulses can produce a symmetry breaking in the thermodynamic limit and in systems with dimension $D>1$. Another recent study\cite{Peronaci2019} based on DMFT, in which the system was driven at a frequency larger than $U$ and at the same time cooled by the coupling to a phonon bath, has found an enhancement of the double occupation, which was interpreted as a possible signature that the system wants to transition into the $\eta$-paired state.  

In a separate work, we have explored the phase diagram of the photo-doped Hubbard model and revealed the existence of $\eta$-pairing over a wide range of parameters.\cite{Li2019} In this case, photo-doped states were prepared in a nonequilibrium steady-state set-up by coupling to particle reservoirs which can insert/remove particles into/from the Hubbard model. The cooling-by-doping protocol provides an alternative way to prepare such states, which works particularly well in the strongly photo-doped regime, with double occupancies close to $d=0.5$. Moreover, with suitably optimized protocols the real-time cooling-by-doping scheme might allow to realize the $\eta$-pairing superconductivity in experiments. In the following, we focus on the analysis of the cooling by doping protocol, while a detailed study of the properties of the $\eta$-pairing state is presented in Ref.~\onlinecite{Li2019}.

We use the symmetric doping scheme to create a large density of doublons and holons in a Mott insulator with $U=9$.  The chirped pulse has the form
\begin{equation}
\Omega(t)=\Omega_\text{min}+(\Omega_\text{max}-\Omega_\text{min})\sin\Big(\frac{\pi}{2}\frac{t}{t_\text{ramp}}\Big),
\end{equation}
with $\Omega_\text{min}$ chosen such that the doublons are initially inserted at the lower edge of the upper Hubbard band, and $\Omega_\text{max}$ large enough that the pulse creates an almost complete population inversion. 

\begin{figure}[t]
\begin{center}
\includegraphics[angle=-90, width=0.8\columnwidth]{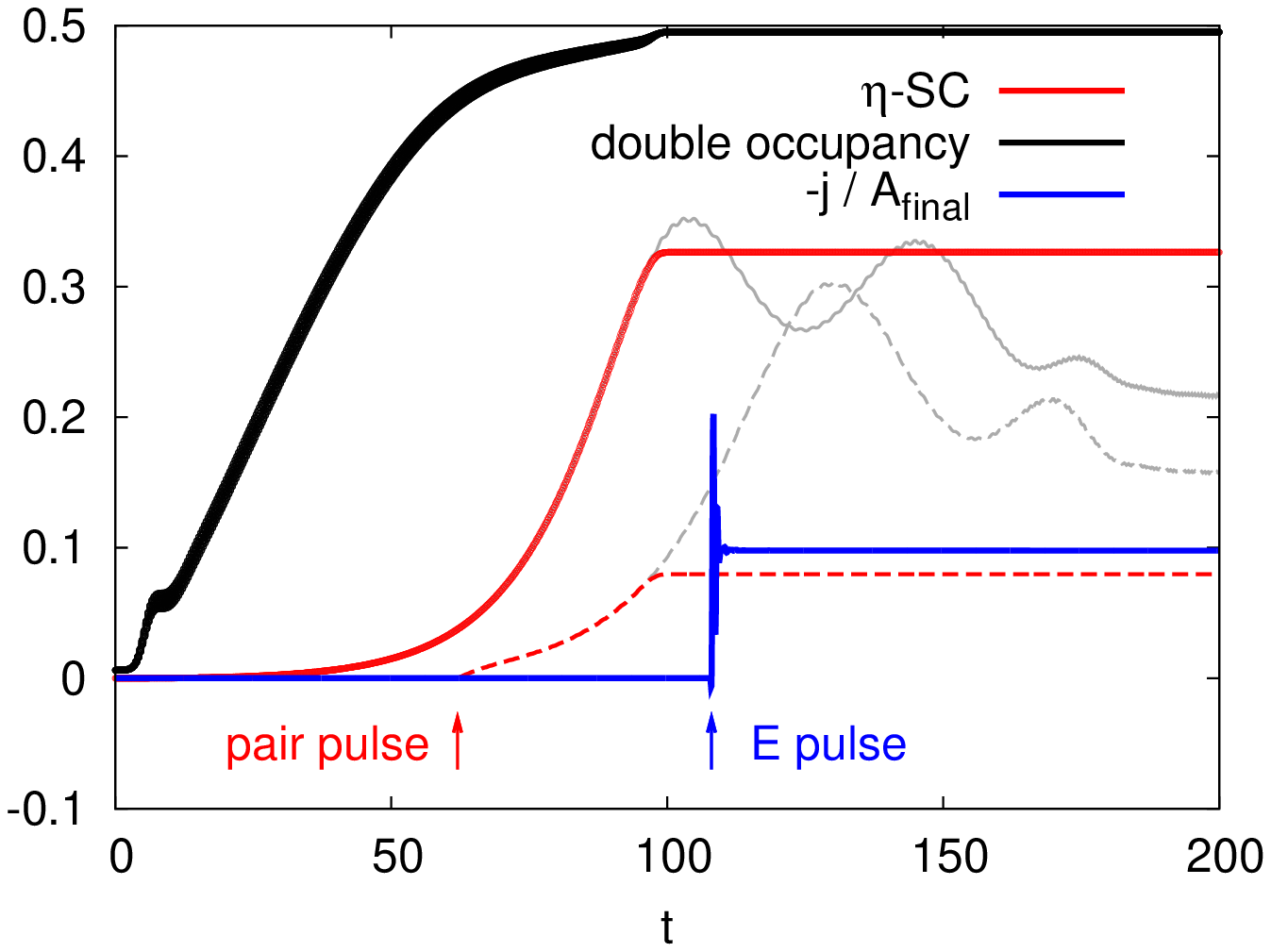}
\includegraphics[angle=-90, width=0.8\columnwidth]{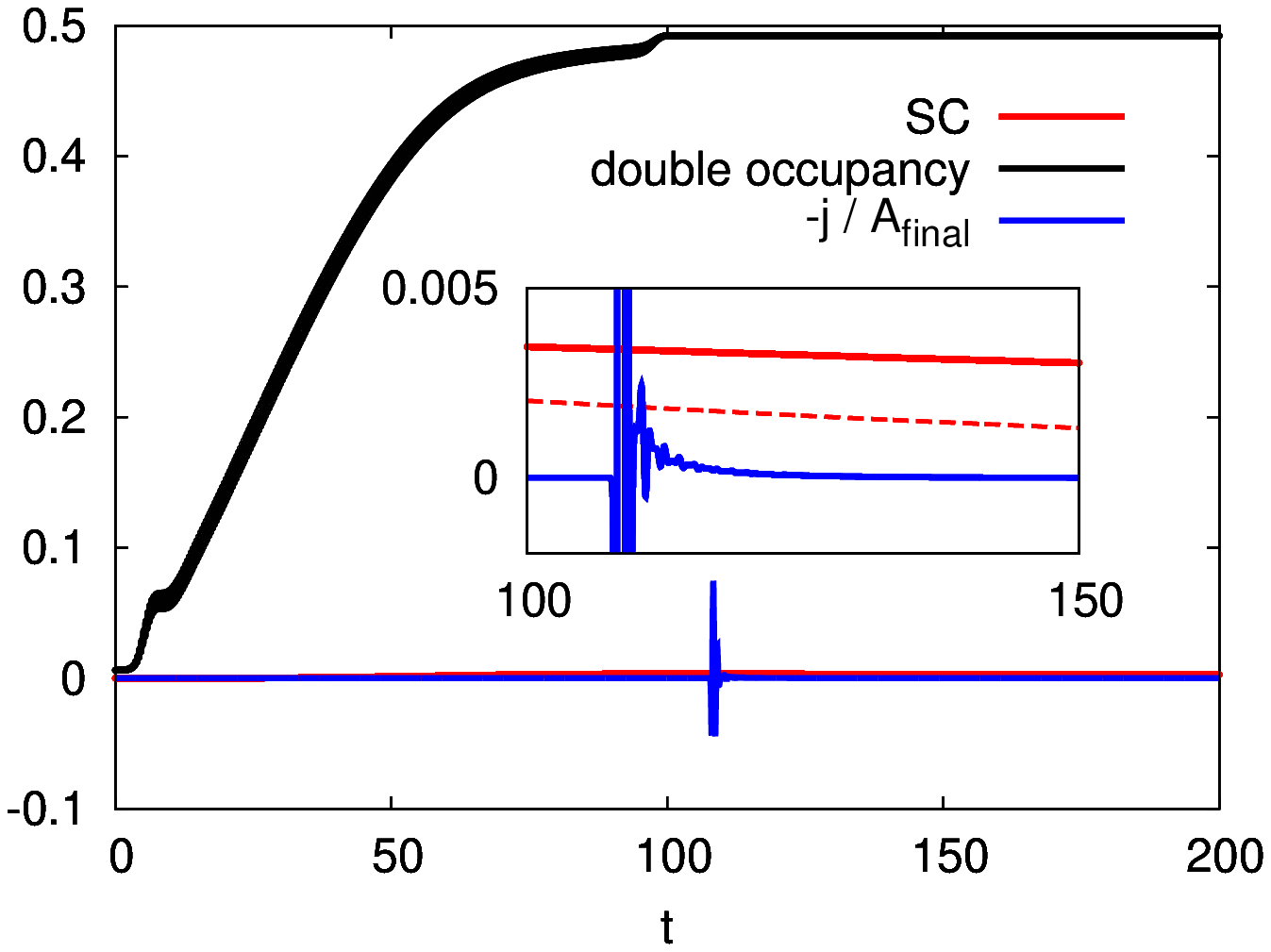}
\caption{Simulation results for the $\eta$-pairing selfconsistency condition (\ref{eta}) (top panel) and the usual $s$-wave pairing selfconsistency (\ref{sc}) (bottom panel) for $U=9$ and initial inverse temperature $\beta=5$, with an applied staggered or uniform pair field of 0.001 (solid lines). 
The red curve shows the superconducting order parameter, the black curve the double occupation, and the blue curve the current induced by a short and weak probe field pulse applied at $t=108$, divided by the vector potential of the field after the pulse. Dashed red lines indicate the order parameter induced by a short and weak pair field pulse at $t=62$, in a simulation without constant pair field. The gray lines in the upper panel show the evolution of the order parameter for pulse duration $>200$. 
}
\label{fig_eta}
\end{center}
\end{figure}

The top panel of Fig.~\ref{fig_eta} plots the time evolution of the double occupation and $\eta$-pairing order parameter for $\Omega_\text{min}=8$, $\Omega_\text{max}=12.5$, $t_\text{ramp}=170$, pulse duration $\approx 100$ and narrow bands of width $0.05$ positioned at energy $\pm 6$. The narrow bands have a box-shaped density of states with Fermi-function like edges corresponding to a cutoff temperature $0.01$. A constant pair field of 0.001 is applied. As the inverted population is building up, the $\eta$-pairing order parameter grows and quickly reaches large values that can only be realized in the spontaneous symmetry-breaking regime (see solid red line). As a direct proof of spontaneous symmetry breaking, we also plot by the dashed red line the order parameter induced by a short and weak pair field pulse at time $t=62$, without constant pair field. Also in this case, the order parameter grows as long as the photo-doping pulse is on. After the switch-off of the photo-doping pulse at $t\approx 100$ the system is decoupled from the narrow bands and the order parameter is conserved under the time evolution. (If the pulse is not stopped, one observes amplitude oscillations in the solid red curve, and a further increase of the dashed red curve, before an eventual melting of the order.) 
Also plotted in the figure is the current induced by a weak half-cycle electric field pulse applied to the system at $t=108$, in the $\eta$-paired state. The current is divided by the vector potential after the pulse, $A_\text{final}=-\int_0^\infty ds E(s) = -0.02$, so that the constant value in the long-time limit represents the delta function contribution to the optical conductivity.  The non-decaying current is another direct proof of the superconducting nature of the photo-induced $\eta$-paired state. 

\begin{figure}[t]
\begin{center}
\includegraphics[angle=-90, width=0.8\columnwidth]{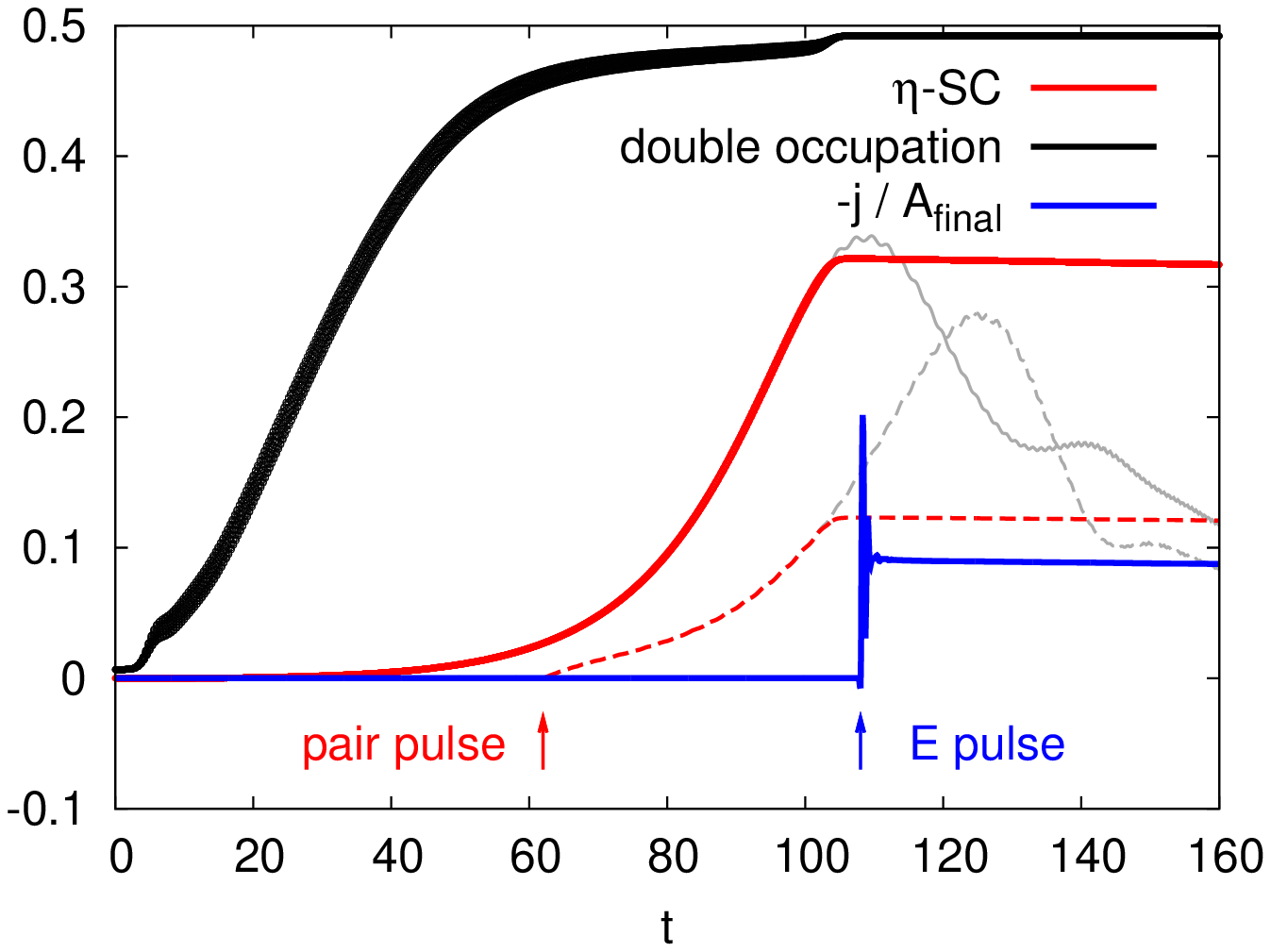}
\includegraphics[angle=-90, width=0.8\columnwidth]{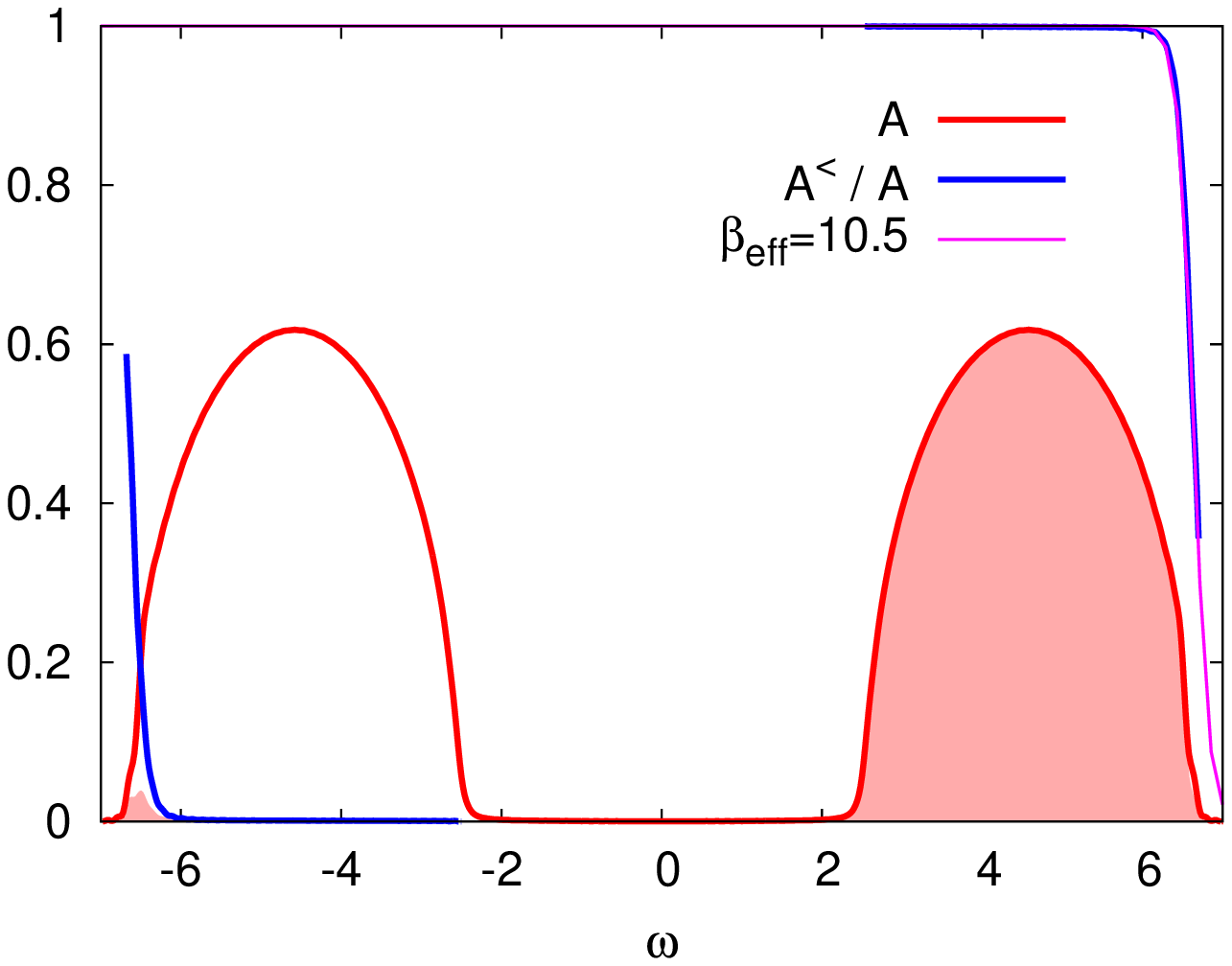}
\caption{Results for the model 
with next-nearest neighbor hopping $v_\text{system}^\text{NNN}=0.25v_\text{system}^\text{NN}$. The top panel shows the double occupation and $\eta$-pairing oder parameter during and after the symmetric cooling-by-photo-doping process. The interaction strength is $U=9$, the initial inverse temperature $\beta=5$, and the applied staggered pair field is 0.001.
The dashed red line indicates the order parameter induced by a weak pair field pulse at $t=62$ in a simulation without constant pair field, and the blue curve the current induced by a short and weak 
probe field pulse applied at $t=108$, divided by the vector potential of the field after the pulse.  
Gray lines indicate the evolution of the order parameter for pulse duration $>160$.  
The bottom panel shows the spectral function and population measured after the end of the pulse ($t=110$), as well as the distribution function $A^</A$ and a fit to a Fermi function in the energy region of the upper Hubbard band.
}
\label{fig_eta_mixed}
\end{center}
\end{figure}

For comparison, we also show in the lower panel the results obtained for the usual $s$-wave superconducting self-consistency loop, for otherwise identical parameters and fields. In this case the order parameter is only slightly enhanced (too little to claim a spontaneous symmetry breaking) and the current induced by the electric field pulse applied to the system is very small and quickly decays back to zero. Also the order parameter in the simulation with the weak pair field pulse (dashed red line) decays back to zero. Therefore, a photo-doped state with a positive effective temperature and a large density of doublons and holons is not susceptible to conventional $s$-wave pairing, but to $\eta$-pairing.\cite{Rosch2008}

We have also simulated a system with next-nearest neighbor hopping equal to one quarter of the nearest-neighbor hopping, $v_\text{system}^\text{NNN}/v_\text{system}^\text{NN}=0.25$. The results for a similar pulse form as in the previous figure are shown in Fig.~\ref{fig_eta_mixed}. An $\eta$-paired state with a comparable magnitude of the order parameter can also be induced in this case, but now the order parameter is no longer conserved after the decoupling of the narrow bands. It decreases slightly at long times, since the narrow bands are decoupled near the maximum of the first ``Higgs" oscillation, and also because of thermalization and heating effects. The gray lines indicate the evolution of the order parameter for pulse duration $>160$. In the simulation with weak pair field pulse (dashed red line), the order parameter also decreases slowly after the switch-off of the pulse, in contrast to the simulation with continuous driving, where it grows up to $t\approx 125$. This indicates that the coupling to the narrow bands is essential for the growth of the order parameter, since it allows to remove the entropy released by the symmetry-breaking from the system.   

Despite the thermalization effects, the system remains in the symmetry-broken state much beyond the longest simulation times. The effective inverse temperature of the doublons and holons measured after the decoupling is approximately $\beta_\text{eff}=10.5$, see bottom panel, which shows the population and energy distribution at $t=110$, including a fit of the energy distribution $A^</A$ to a Fermi function (pink curve).   

Note that the $\eta$-paired state is a long-lived nonequilibrium state of the system, which in contrast to the negative-temperature state is not expected to last forever in the isolated system. On exponentially long timescales, doublon-holon recombination and scattering processes will thermalize the system and lead to a melting of the order. (Because of the large energy of the population inverted state, this thermalized state will be a negative temperature state.) On the other hand, due to the positive $\beta_\text{eff}$, the transient state realized after the photo-doping is robust against cooling by phonons or other baths with positive temperature, in contrast to the conventional superconducting state with negative $\beta_\text{eff}$ discussed in Sec.~\ref{sec:sc}, which is not expected to last very long in the presence of energy dissipation.

\section{Conclusions}
\label{sec:conclusions}

We have demonstrated the versatility of the cooling-by-doping approach by producing different types of nonequilibrium states of the Hubbard model, which were inaccessible in previous nonequilibrium DMFT based studies. The formation of photo-doped Mott insulating states with low effective temperature have been hampered by the high entropy and high effective temperature resulting from doublon-holon production in an initially Mott insulating single-band system, while the formation of quasi-particle bands in a photo-doped state cooled by the coupling to a boson bath was found to be very slow.  Here, we showed that the simultaneous doping of the Hubbard bands by particle exchange with narrow bands and the transfer of entropy into these bands allows to realize cold photo-doped states with sharp quasi-particle features in the spectral function and optical conductivity. We also showed that the optical conductivity of this state is essentially identical to that of a chemically doped state with the same temperature and a doping concentration which amounts to the photo-doped density of doublons and holons. It appears that cooling-by-doping circumvents potential bottlenecks in the formation of quasi-particles by evolving the system along the filling, rather than temperature axis, and by avoiding the passage through high-entropy regions.  

Using a similar doping protocol, we also realized a negative-temperature state in a moderately correlated Hubbard model, which was cold enough for the symmetry-breaking into a superconducting state. This superconducting state is equivalent to the usual $s$-wave superconducting state realized in the attractive Hubbard model, since there is an exact mapping between the repulsive system with $\beta_\text{eff}<0$ and the attractive system with $\beta_\text{eff}>0$. While it is intriguing to think about the possibility of realizing such a superconducting state in photo-excited materials, this state would probably not survive for a long time, since a coupling to phonons or other baths with $\beta>0$ would destabilize the negative temperature distribution in the photo-excited system. 

More intriguing from a practical point of view is the possibility of producing an $\eta$-paired state in a strongly interacting system with a large density of doublons and holons. We have shown that such a state can in principle be prepared by a short electric field pulse using the cooling-by-doping scheme. While this state is expected to decay due to doublon-holon recombination and associated heating, the life-time depends exponentially on the interaction strength or gap size, and can in practice be very long.  

The present work presents a proof of principles, and demonstrates a powerful theoretical tool for the preparation and study of nonequilibrium states. While the set-up considered in this study may look artificial, it will be worthwhile to also explore the possible role of cooling-by-doping in experiments. The effect is strong and based on a very simple physical principle -- evaporative cooling -- which can be exploited both in condensed matter and cold atom contexts. 
In a cold atom set-up, the coupling of different bands by hopping modulation is the most natural procedure, so that our set-up is directly relevant for these experiments. The threshold for antiferromagnetic order has only very recently been reached in cold atom systems using a different entropy cooling scheme,\cite{Mazurenko2017} and our insights may provide a path for realizing even lower temperatures and more exotic phases. In this case, the transfer of particles into empty narrow bands may be the most natural path. The bandstructure of the brickwall lattice,\cite{Taruell2012} which has been used in recent lattice shaking experiments,\cite{Sandholzer2018} features such a flat empty band. In fact, this lattice has been mainly implemented because the flat band allows to suppress unwanted excitations. In view of our results, it would also be very interesting to experimentally explore the cooling effect resulting from resonant excitations of atoms into the flat empty band. More generally, cooling-by-doping should affect any experimental protocol which involves a substantial particle transfer between different bands, and in particular particle transfer out of narrow full bands, or into narrow empty bands. 

\acknowledgements{
The calculations were run on the Beo04 cluster at the University of Fribourg, using a software library co-developed by H. Strand. PW was supported by the European Research Council through ERC Consolidator Grant 724103. M.E. and J. Li were supported by ERC Starting Grant No. 716648. The Flatiron institute is a division of the Simons Foundation. PW thanks the Aspen Center for Physics, which is supported by National Science Foundation grant PHY-1607611, for its hospitality during the summer 2019 program on ``Realizations and Applications of Quantum Coherence in Non-Equilibrium Systems."
} 

\appendix

\section{Pulse parameters}
\label{app}

For the sake of data reproducibility, we list here the precise pulse shapes used in the simulations:

\begin{itemize}
\item Fig.~2, left panels: $a_\text{max}=0.15$, $f_\text{envelope}(t)=1/[(1+e^{t-50+4})(1+e^{-(t-4)})]$, $\Omega(t)=8.5+(8.88475-8.5)\sin(\tfrac{t}{50}\tfrac{\pi}{2})$, probe electric field pulse: $E(t)=1/[(1+e^{20(t-60+0.04)})(1+e^{-20(t-60-0.04)})]$.
\item Fig.~2, middle panels: $a_\text{max}=0.25$, $f_\text{envelope}(t)=1/[(1+e^{t-50+4})(1+e^{-(t-4)})]$, $\Omega(t)=8.5+(9.2545-8.5)\sin(\tfrac{t}{50}\tfrac{\pi}{2})$, probe electric field pulse: $E(t)=1/[(1+e^{20(t-60+0.04)})(1+e^{-20(t-60-0.04)})]$.
\item Fig.~2, right panels: $a_\text{max}=0.25$, $f_\text{envelope}(t)=1/[(1+e^{t-50+4})(1+e^{-(t-4)})]$, $\Omega(t)=8.5+(9.5282-8.5)\sin(\tfrac{t}{50}\tfrac{\pi}{2})$, probe electric field pulse: $E(t)=1/[(1+e^{20(t-60+0.04)})(1+e^{-20(t-60-0.04)})]$.
\item Fig.~5: $a_\text{max}=0.45$, $f_\text{envelope}(t)=1/[(1+e^{t-150+6})(1+e^{-(t-4)})]$, $\Omega(t)=10+(7.5-10)\tfrac{t}{80}$ if $t \le 62.5$,  $\Omega(t)=10+(7.5-10)2(\tfrac{62.5}{80})$ (plus phase shift) if $t > 62.5$.
\item Fig.~6: $a_\text{max}=0.65$, $f_\text{envelope}(t)=1/[(1+e^{t-102+4})(1+e^{-(t-4)})]$, $\Omega(t)=8+(12.5-8)\sin(\tfrac{t}{170}\tfrac{\pi}{2})$,
probe electric field pulse: $E(t)=1/[(1+e^{20(t-108+0.04)})(1+e^{-20(t-108-0.04)})]$,
pair field pulse: $p(t)=1/[(1+e^{20(t-62+0.04)})(1+e^{-20(t-62-0.04)})]$.
\item Fig.~7: $a_\text{max}=0.85$, $f_\text{envelope}(t)=1/[(1+e^{t-108+4})(1+e^{-(t-4)})]$, $\Omega(t)=7.25+(12.5-7.25)\sin(\tfrac{t}{140}\tfrac{\pi}{2})$,
probe electric field pulse: $E(t)=1/[(1+e^{20(t-108+0.04)})(1+e^{-20(t-108-0.04)})]$,
pair field pulse: $p(t)=1/[(1+e^{20(t-62+0.04)})(1+e^{-20(t-62-0.04)})]$.
\end{itemize}

\end{document}